\documentclass[11pt,twocolumn]{article}
\usepackage[dvips]{graphicx}
\usepackage{vmargin}  
    \setpapersize{A4}
    \setmarginsrb{1.3cm}{2cm}{2cm}{2.5cm}{0pt}{0mm}{0pt}{0mm}
\usepackage{newcent}  
\usepackage{natbib}
     \bibpunct{[}{]}{;}{a}{,}{,}

\begin{document}

\title{Simple models for scaling in phylogenetic trees}
\author{Emilio Hern\'andez-Garc{\'\i}a$^1$, Murat Tu\u{g}rul$^1$, E. Alejandro
Herrada$^1$, \\
V{\'\i}ctor M. Egu{\'\i}luz$^1$, Konstantin
Klemm$^{1,2}$\\\
$^1$IFISC (UIB-CSIC) Instituto de F{\'\i}sica Interdisciplinar y
Sistemas Complejos,  \\
Campus Universitat de les Illes Balears, E-07122 Palma de
Mallorca, Spain\\
$^2$Bioinformatics, Department of Computer Science, University
Leipzig\\ H{\"a}rtelstr. 16-18, 04107 Leipzig, Germany}

\date{January 30, 2009}

\maketitle

\begin{abstract}
Many processes and models --in biological, physical, social, and
other contexts-- produce trees whose depth scales logarithmically
with the number of leaves. Phylogenetic trees, describing the
evolutionary relationships between biological species, are
examples of trees for which such scaling is not observed. With
this motivation, we analyze numerically two branching models
leading to non-logarithmic scaling of the depth with the number of
leaves. For Ford's alpha model, although a power-law scaling of
the depth with tree size was established analytically, our
numerical results illustrate that the asymptotic regime is
approached only at very large tree sizes. We  introduce here a new
model, the {\sl activity} model, showing analytically and
numerically that it also displays a power-law scaling of the depth
with tree size at a critical parameter value.
\end{abstract}


\section{Phylogenetic branching and models}

Although most modern studies on complex networks
\citep{Albert2002,Boccaletti2006} consider situations in which
nodes are connected by multiple paths, the case of {\sl trees},
i.e. graphs without closed cycles, is relevant to describe many
natural and artificial systems. Branching in real trees
\citep{Stevens1974}, in blood vessels \citep{West1997}, in river
networks \citep{RodriguezIturbe1997} or in computer file systems
\citep{Klemm2005,Klemm2006} produce complex tree patterns worth to
be described and understood. Trees are the outcome of
classifications algorithms \citep{Jain1988} and of branching
processes \citep{Harris1963} and they also arise when computing
community structure \citep{Guimera2003} or as a backbone (for
example a minimum spanning tree) for more connected networks
\citep{Garlaschelli2003,HernandezGarcia2007,Rozenfeld2008}.

Evolutionary processes leading to speciation are also summarized
in phylogenetic trees \citep{Cracraft2004}. In these trees the
leaves represent living species and each internal node represents
a branching event in which an ancestral species diversified into
daughter species. Every internal node is thus the root of its
associated subtree which consists of all its descendant nodes.
Phylogenetic tree topology encodes information on evolutionary
mechanisms which is beginning to be scrutinized
\citep{Burlando1990,Burlando1993,Ford2006,Blum2006,HernandezGarcia2007,Herrada2008}.

\begin{figure}[ht]
    \begin{center}
    \includegraphics[width=.9\columnwidth,clip]{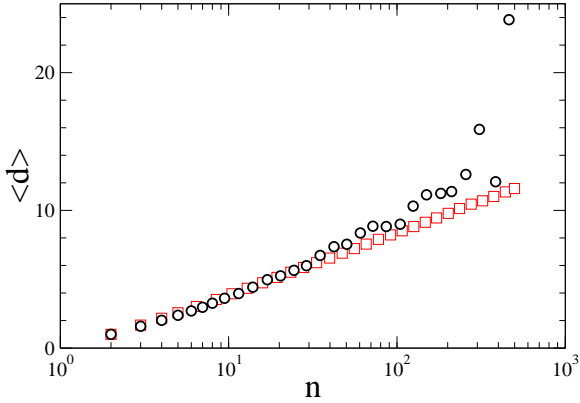}
    \end{center}
\caption{\label{fig:data} Mean depth $\left<d\right>$ of trees in
TreeBASE (circles) as a function of number of leaves $n$. Squares
are obtained from computer simulations of the ERM model, behaving
as Eq. (\ref{log}) for large $n$. At large sizes, the depth in the
real phylogenetic trees scales with the number of leaves faster
than the ERM behavior. For both real phylogenies and model, depth
values for each tree size are obtained by logarithmic binning of
the depth of all trees and subtrees with that size.}
\end{figure}

\begin{figure}[hbt]
    \begin{center}
    \includegraphics[width=1.\columnwidth,clip]{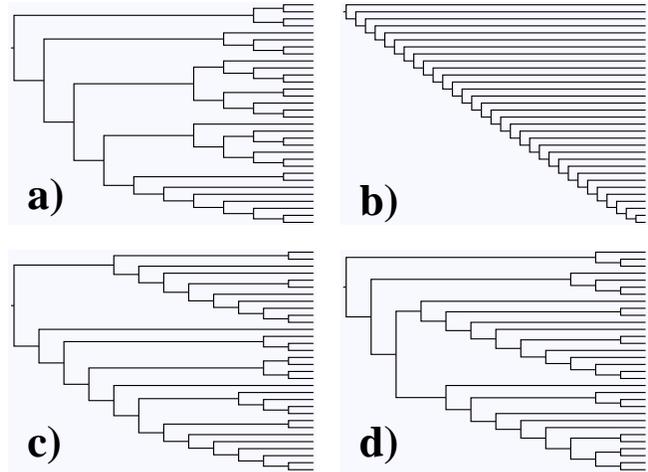}
    \end{center}
\caption{\label{fig:examples} Examples of trees with 32 leaves, generated from
several models. a) Tree generated with the ERM model, which is
equivalent to the alpha model with $\alpha=0$. b) The completely
unbalanced tree, which is equivalent to the alpha model with
$\alpha=1$. c) A tree generated with the alpha model for
$\alpha=0.5$. d) A tree generated with the activity model for
$p=0.5$. The trees in c) and d) display an imbalance intermediate
between a) and b).}
\end{figure}

The earliest mathematical model of evolutionary branching was
proposed by \citet{Yule1925}. Apart from the distinction he
introduced between genera and species diversification, the model
is equivalent to the Equal Rates Markov (ERM) model
\citep{Harding1971,Cavalli1967}: starting from a single ancestral
species, one among the tree leaves existing at the present time is
chosen at random, bifurcating into two new leaves. Then this
operation is repeated for a number of time steps or, equivalently,
until the tree reaches a desired size. The topological
characteristics of the constructed trees are surprisingly robust,
being shared by apparently different models such as the coalescent
and others \citep{Aldous2001}. Essentially what is needed is that
different branches at a given time branch independently and with
the same probabilities. When extinction is taken into account, the
same topology is recovered when considering only the lineages
surviving at the final time. One of the characteristics of this
type of branching is a distribution of subtree sizes $A$ scaling
at large sizes as $A^{-2}$, an outcome robustly observed in many
natural and artificial systems and in classification schemes,
including taxonomies
\citep{Burlando1990,Caldarelli2004,Capocci2008}. Another important
characteristic is that the mean depth of the tree $\left<d\right>$
(i.e. the average distance, measured in number of links, from the
leaves to the root) scales logarithmically with the number of
leaves $n$:
\begin{equation}
\left<d\right>\sim \log n \ .
\label{log}
\end{equation}
It is worth noting that these results apply not only to many
random branching models, but also to the simple deterministic
Cayley tree, in which all internal nodes at a given level split in
a fixed number of daughter nodes.

In view of this generality it was surprising to find that the
topology of observed phylogenies does not agree with any of these
predictions \citep{Herrada2008}. In fact, it was known since some
time ago that real phylogenies are substantially more {\sl
unbalanced} than predicted by the ERM and similar models
\citep{Aldous2001,Blum2006}. This means that some lineages
diversify much more than others, in a way that is statistically
incompatible with the ERM predictions. Figure \ref{fig:data}
compares data \citep{Herrada2008} compiled from TreeBASE, a public
repository containing several thousands of empirical phylogenetic
trees corresponding to virtually all kinds of organisms in Earth,
with the predictions of the ERM model. For the phylogenetic trees
at large sizes the mean depth scales with the number of leaves
faster than the ERM behavior in Eq.(\ref{log}).

The breakdown of the ERM behavior indicates that evolutionary
branching should present correlations either in time or between
the different branches. Mechanisms producing trees with non-ERM
scaling for the depth have been identified, as for example the
situation of critical branching \citep{DeLosRios2001,Harris1963}
or optimization of transport processes \citep{Banavar1999}. In the
phylogenetic context models of this type have been proposed
\citep{Aldous2001,Pinelis2003,Blum2006,Ford2006}, although most of
them lack a clear interpretation in biological terms.

In the following we present results for two branching models
showing asymptotically non-ERM, i.e. non-logarithmic, scaling for
the depth. Their study is motivated, on the one hand, by the
empirical results above from real phylogenetic trees. On the
other, they pertain to the small set of available models with
non-ERM scaling which are defined {\sl dynamically} (i.e. by a set
of rules that are applied to the present state of a growing tree
to find the state at the next time step) rather than being
characterized globally by statistical or optimization
prescriptions. The first model we present, Ford's {\sl alpha}
model, is a simple example for which the non-trivial asymptotic
scaling (of the power law type) has been analytically identified.
We analyze it numerically to confirm this prediction and to
display the behavior at finite sizes. We introduce later a new
model, the {\sl activity} model, which also leads to
non-logarithmic depth scaling at a critical parameter value.

\section{Ford's alpha model}

\begin{figure*}[hbt]
    \begin{center}
    \includegraphics[angle=0,width=1.5\columnwidth]{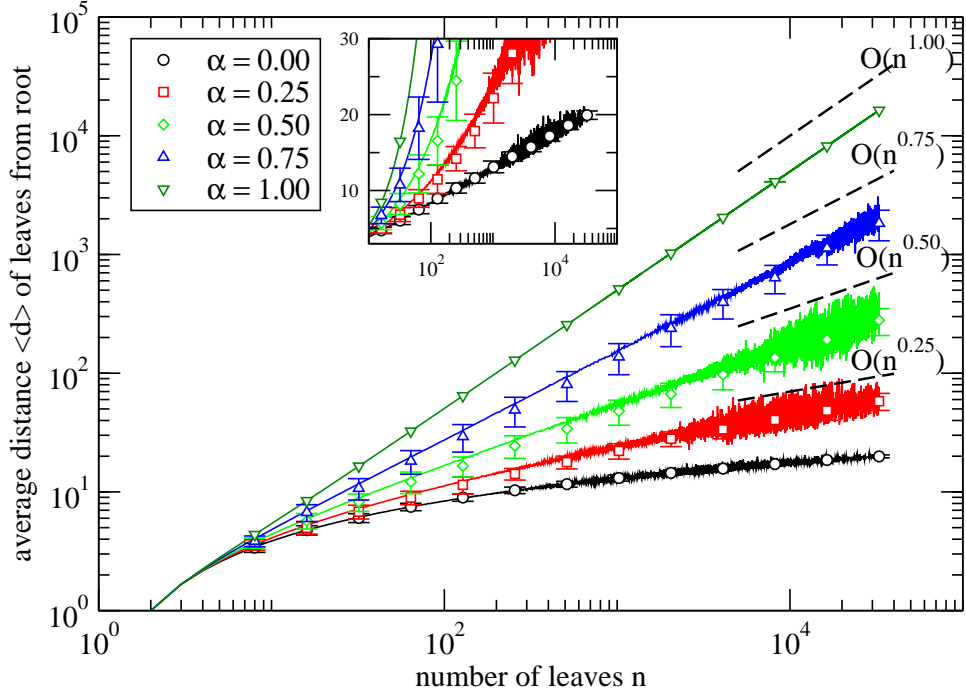}
    \end{center}
\caption{\label{fig:alpha} Depth statistics vs tree size for the alpha model.
Symbols indicate the mean depth of leaves from root, averaged over
the 100 trees generated for each size ($2^k$, $k=3, 4, ..., 15$),
and the error bars are the corresponding standard deviations. The
points in the rugged lines come from each subtree of all trees
generated. The dashed segments indicate the analytic predictions
\citep{Ford2006} for the scaling at large $n$. The inset
highlights the logarithmic scaling of the $\alpha=0$ case.}
\end{figure*}

\citet{Ford2006} introduced a model for recursive tree formation:
At a given step in the process the tree is a set of leaves
connected by terminal links to internal nodes, which are
themselves connected by internal edges until reaching the root
(the root itself is considered to have a single edge, which we
count as internal, joining to the first bifurcating internal node;
with this convention a tree of $n$ leaves has $n-1$ internal
edges). Then, a probability of branching proportional to
$1-\alpha$ is assigned to each leaf, and proportional to $\alpha$
to each internal edge. By normalization these probabilities are,
respectively, $(1-\alpha)/(n-\alpha)$, and $\alpha/(n-\alpha)$.
When a leaf is selected for branching, it gives birth to a couple
of new ones, as in the ERM model. But when choosing an internal
edge, a new leaf branches from it by the insertion in the edge of
a new internal node. For $\alpha=0$ we have the standard ERM
model. For $\alpha=1$ the completely unbalanced comb tree, in
which all leaves branch successively from a main branch, is
generated. Intermediate topologies are obtained for $\alpha \in
(0,1)$. Figure \ref{fig:examples} shows examples of trees
generated for different values of $\alpha$.

By considering the effect of the addition of new leaves on the
distances between root and other nodes, \citet{Ford2006} derived
exact recurrence relationships which, when written in terms of the
average depth, lead to:
\begin{equation}
\left<d\right>_{n+1}=\frac{n}{n-\alpha}\left<d\right>_n +
\frac{2n(1-2\alpha)}{(n+1)(n-\alpha)} \ .
\label{alphadn}
\end{equation}
$\left<d\right>_n$ is the mean depth of the leaves of a tree with
$n$ leaves. By assuming a behavior $\left<d\right>_n \sim n^\nu$
at large $n$, and expanding Eq. (\ref{alphadn}) in powers of
$1/n$, we get $\nu=\alpha$, so that
\begin{equation}
\left<d\right>_n \sim n^\alpha \ , {\rm if }\ \  0<\alpha\le 1 \ .
\end{equation}
If $\alpha=0$ the standard ERM behavior, Eq. (\ref{log}), is
recovered.

Figure \ref{fig:alpha} shows numerical results for the depth of
trees generated with this model. Note that the predicted
asymptotic behavior is attained but only at very large tree sizes,
in general sizes much larger than the tree sizes of the examples
shown in Fig. \ref{fig:examples} and of the available empirical
phylogenies. As analytically demonstrated \citep{Ford2006} depth
statistics of subtrees of given size extracted from a large tree
behave as data from trees of that size directly generated by the
alpha model algorithm.

While the Ford model gives a simple mechanism for scaling in trees
with a tunable exponent, the dynamical rule of posterior insertion
of inner nodes is hard to justify in the context of evolution
(although one can think on the modelling of errors arising in
phylogenetic reconstruction methods when incorrectly assigning a
splitting to a non-existing ancestral species). This motivates the
introduction of a new model described in the next section.

\section{Activity model}

\begin{figure*}[ht]
    \begin{center}
    \includegraphics[width=1.5\columnwidth]{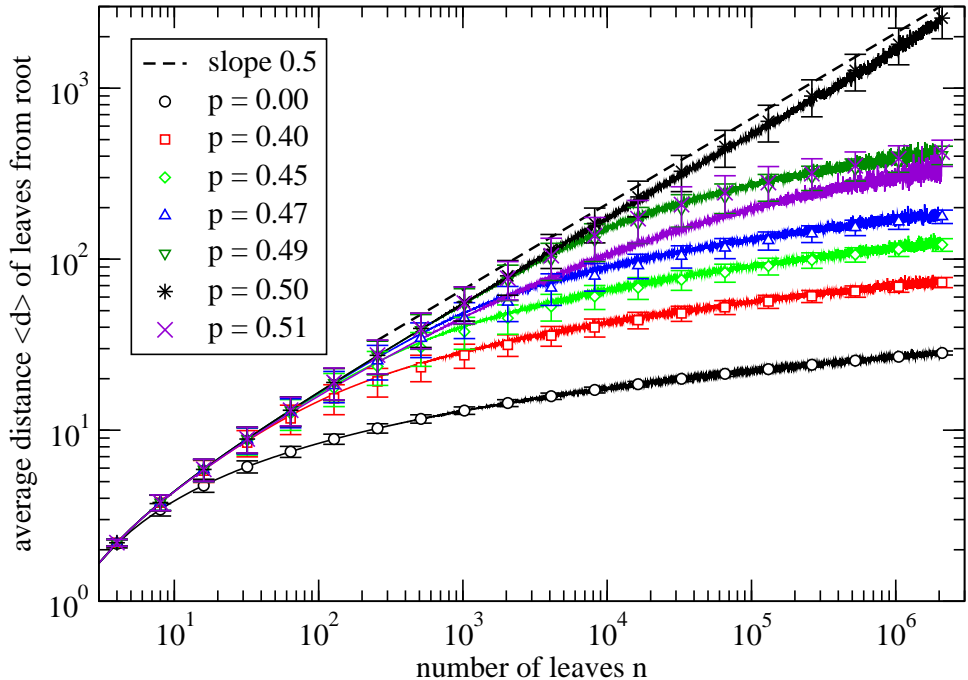}
    \end{center}
\caption{\label{fig:activity} Average depth versus size for the
activity model for various values of the activation probability
$p$. Data points displayed by symbols give the average distance of
leaves with respect to the root. Error bars give the standard
deviation taken over different realizations (1000 trees per data
point). Data in the rugged curves are for all subtrees of trees
with size $2^{21}=2097152$. The dashed line represents a power law
scaling with exponent $1/2$, corresponding to the scaling of the
$p=0.5$ curve, as discussed in the text. }
\end{figure*}

In this section we show that tree shapes distinct from the ERM
model may also result from a memory in terms of internal states of
the nodes. The {\em activity} model proposed here is conceptually
similar to the class of models suggested by \citet{Pinelis2003}.
However, the present model distinguishes only between active and
inactive nodes and has a single parameter controlling the spread
of activity.

Starting from a single node (the root), a binary tree is generated
as follows. At each step, a leaf $i$ of the tree is chosen and
branched into two new leaves. Each of the two new leaves,
independently of the other, is set active with probability $p$ or
inactive with probability $1 - p$. The branching leaf $i$ is
chosen at random from the set of active leaves if this set is
non-empty. Otherwise, $i$ is chosen at random from the set of all
leaves. Figure \ref{fig:activity} shows that for $p = 1/2$ the
model generates trees with mean depth growing as the square root
of tree size (note the log-log scale). Figure \ref{fig:examples}
displays a small-size example of such trees. For values of $p$
below or above $1/2$, $\left<d\right>$ seems to increase
logarithmically with $n$.

Here we give a simplified argument to understand the observed
exponent $1/2$ of the distance scaling with system size in the
case $p = 1/2$. At the time the growing tree has $n$ leaves in
total, let $D_a(n)$ be the expected sum of distances of active
leaves from the root, and $D_b(n)$ the analogous quantity for the
inactive leaves. When a randomly chosen active leaf --at distance
$d_a$ from root-- branches, the expected increase of $D_a(n)$ is
\begin{eqnarray}
\Delta D_a(n) & \equiv & D_a (n+1) - D_a (n) = \nonumber \\
p^2 (d_a+2) &+& 2p(1-p)\cdot 1 + (1-p)^2 (-d_a) \nonumber \\
&=& (2p-1)d_a+2p \ .
\end{eqnarray}
Here the three terms of the second line are for the activation of
two, one and zero of the new leaves, respectively. This expression
is appropriate as far as the number of active nodes is not zero.
Simultaneously, the expected change in $D_b(n)$ during the same
event is
\begin{eqnarray}
\Delta D_b(n) &=& \nonumber \\
p^2 \cdot 0 &+& 2p(1-p) (d_a+1)+ (1-p)^2 2(d_a+1) \nonumber
\\ &=& 2(1-p) (d_a+1) \ .
\end{eqnarray}
We now average $\Delta D_a(n)$ over the different choices of the
particular active leave that has been branched. This amounts to
replacing $d_a$ in the above formulae by $\left<d_a\right>_n$, the
average depth of the {\sl active} leaves in a tree of $n$ leaves.
Writing $D_i(n+1)=D_i(n)+\Delta D_i(n)$, for $i=a,b$, one would
get a closed system for the quantities $D_i(n)$ provided
$\left<d_a\right>_n$ is expressed in terms of them. This can be
done by writing $\left<d_a\right>_n=D_a(n)/a(n)$, where $a(n)$ is
the expected number of active leaves in a tree of $n$ leaves. This
expected value is used here as an approximation to the actual
number of active leaves.

The recurrence equations for $D_i(n)$ are specially simple in the
most interesting case $p=1/2$, since the dependence in
$\left<d_a\right>_n$ disappears from one of the equations:
\begin{eqnarray}
D_a(n+1) &=& D_a(n)+1 \label{activitya}\\
D_b(n+1) &=& D_b(n)+\left<d_a\right>_n+1 \ .
\label{activityb}
\end{eqnarray}
The solution (with initial condition $D_a(1)=0$) of Eq.
(\ref{activitya}) is simply:
\begin{equation}
D_a(n) = n-1 \ .
\label{activityDan}
\end{equation}
Since the probabilities of an increment or decrement (by one unit)
of the number of active leaves are the same and time-independent
for $p=1/2$, the number of active nodes performs a symmetric
random walk with a reflecting boundary at 0 (this last condition
arises from the prescription of setting active one node when the
number of active nodes has reached zero in the previous step). For
such random walk the expected value of active leaves $a(n)$
increases as the square root of the number of steps. Since a new
leaf is added at each time step, this leads to:
\begin{equation}
a(n) \sim n^{1/2} \ .
\label{activityan}
\end{equation}
Combining (\ref{activityDan}) and (\ref{activityan}) we obtain the
average distance of active nodes from root at large tree sizes:
\begin{equation}
\left<d_a\right>_n \approx \frac{D_a(n)}{a(n)} \sim n^{1/2} \ .
\end{equation}

Now we can plug this result into Eq. (\ref{activityb}), which can
be solved recursively:
\begin{equation}
D_b(n)=D_b(1)+\sum_{t=1}^{n-1} \left(  \left<d_a\right>_t +1
\right) \sim \sum_{t=1}^{n-1} t^{1/2} \sim n^{3/2} \ .
\label{activityDbn}
\end{equation}
The totally averaged depth $\left< d\right>_n$, which counts both
the active and the inactive leaves, is
\begin{equation}
\left<d\right>_n = \frac{D_a(n) + D_b(n)}{n} \sim \frac{n^{1/2} +
n^{3/2}}{n} \sim n^{1/2} \ ,
\end{equation}
which explains the asymptotic behavior observed in Fig.
\ref{fig:activity} for $p=1/2$.

We note that the growth dynamics presented here may be mapped to a
branching process \citep{Harris1963}, with the difference that
here the death (inactivation) of a node does not lead to its
removal from the tree. The special case $p=1/2$ corresponds to a
critical branching process.

\section{Discussion}

We have presented and studied two simple models which lead to
non-logarithmic scaling of the tree depth. In contrast with many
of the available models having this behavior
\citep{Banavar1999,Aldous2001,Blum2006,Ford2006} they are
formulated as {\sl dynamical} models involving {\sl growing
trees}, so that rules are given to obtain the tree at the next
time step from the present state. Their study has been motivated
by data from phylogenetic branching, and they are interesting
additions to our present understanding of complex networks and
trees.

A recent analysis of several evolutionary models including species
competition \citep{Stich2008} indicates that in these models
correlations are finally destroyed by mutation processes and
persist only for a finite correlation time. Thus sufficiently
large trees would have a scaling behavior closer to the asymptotic
ERM predictions. Since the largest phylogenies in databases such
as TreeBASE have only some hundreds of leaves, it is possible that
the observed imbalance and depth scaling is a finite-size regime.
Nevertheless models going beyond the ERM scaling are needed at
least to explain this finite-size regime, and also to elucidate
the true asymptotic scaling behavior. Here, we have also observed
large finite-size transients in the alpha model of Sect. 2.

The different types of scaling of depth with size can be
interpreted as indicating different values of the (fractal)
dimensionality of the trees. This is so because $\left<d\right>$
is a measure of the {\sl diameter of the tree}, and because for a
binary tree the total number of nodes is simply twice the number
of leaves. Since the simplest definition of dimension $D$ of a
network \citep{Eguiluz2003} is given by the growth of the number
of nodes as the diameter increases, $n \sim \left<d\right>^D$,
power law scaling of the type $\left<d\right> \sim n^\nu$
indicates that the tree can be thought as having a dimension
$D=1/\nu$. The logarithmic scaling in the ERM model is an example
of the {\sl small-world} behavior common to many network
structures \citep{Albert2002}, which is equivalent to having an
effective infinite dimensionality, whereas the power law scaling
reveals a finite dimension for the tree, which implies a more
constrained mode of branching. The alpha model produces trees with
tunable dimension from 1 to $\infty$, and the critical activity
model gives two-dimensional trees.

The final aim of the modelling of phylogenetic trees is to provide
biological mechanisms explaining the branching topology of the
Tree of Life. In this direction, the branching of internal edges
in the Ford model has no obvious biological interpretation. The
activity model puts the mechanisms of birth-death critical
branching \citep{Harris1963} within a framework of transitions
between node internal states similar in spirit to the approach of
\citet{Pinelis2003}. The need to tune a parameter to attain the
non-ERM critical behavior is however a limitation for its
applicability. Much additional work is needed to identify the
proper biological mechanisms behind evolutionary branching and
adequate modelling of them.

\section*{Acknowledgments}

We acknowledge financial support from the European Commission
through the NEST-Complexity project EDEN (043251) and from MICINN
(Spain) and FEDER through project FISICOS (FIS2007-60327).



\end{document}